\documentclass[conference]{IEEEtran}
\IEEEoverridecommandlockouts

\usepackage[T1]{fontenc}
\usepackage[brazil,english]{babel}
\usepackage{amsmath,amsfonts,amssymb}
\usepackage{algorithm}
\usepackage{pifont}
\usepackage{graphicx}
\usepackage{subfigure} 
\usepackage{psfrag}

\usepackage{enumitem}
\usepackage{array}
\usepackage{units}
\usepackage{booktabs} 
\usepackage{lineno}
\usepackage{steinmetz} 
\usepackage{multirow}
\usepackage{xcolor}
\usepackage{cite}
\usepackage{array}
\usepackage{url}
\makeatletter
\def\url@leostyle{%
  \@ifundefined{selectfont}{\def\UrlFont{\sf}}{\def\UrlFont{\small\ttfamily}}}
\makeatother
\usepackage{hyperref}
\usepackage{breakurl}

\begin{document}

\title{Performance Comparison of Deep RL Algorithms for Energy Systems Optimal Scheduling\\

\thanks{978-1-6654-8032-1/22/\$31.00 ©2022 European Union}
}

\author{\IEEEauthorblockN
{$\text{Hou Shengren}^{1}$, $\text{Edgar Mauricio Salazar}^{2}$, $\text{Pedro P. Vergara}^{1}$, $\text{Peter Palensky}^{1}$}
\IEEEauthorblockA{${^1}\text{Intelligent Electrical Power Grids, Delft University of Technology, The Netherlands}$} 
\IEEEauthorblockA{${^2}\text{Electrical Energy Systems (EES) Group, Eindhoven University of Technology, The Netherlands}$ \\
emails: \{H.Shengren, P.P.VergaraBarrios, P.Palensky\}@tudelft.nl,  E.M.Salazar.Duque@tue.nl}
}
\maketitle

\begin{abstract}
Taking advantage of their data-driven and model-free features, Deep Reinforcement Learning (DRL) algorithms have the potential to deal with the increasing level of uncertainty due to the introduction of renewable-based generation. To deal simultaneously with the energy systems' operational cost and technical constraints (e.g, generation-demand power balance) DRL algorithms must consider a trade-off when designing the reward function. This trade-off introduces extra hyperparameters that impact the DRL algorithms' performance and capability of providing feasible solutions. In this paper, a performance comparison of different DRL algorithms, including DDPG, TD3, SAC, and PPO, are presented. We aim to provide a fair comparison of these DRL algorithms for energy systems optimal scheduling problems. Results show DRL algorithms' capability of providing in real-time good-quality solutions, even in unseen operational scenarios, when compared with a mathematical programming model of the energy system optimal scheduling problem. Nevertheless, in the case of large peak consumption, these algorithms failed to provide feasible solutions, which can impede their practical implementation. 
\end{abstract}

\section*{Notation} \vspace{-2mm}
\noindent \emph{Sets}:
\begin{description}[itemsep=0.5ex, labelwidth=1.2cm, leftmargin=1.4cm]
	\item [{\small${\cal G}$}]      Set of (DGs) distributed generators
	\item[{\small$\cal{B}$}] Set of ESSs
	\item[{\small$\cal{L}$}] Set of Loads
	\item[{\small$\cal{V}$}] Set of PVs 
	\item [{\small${\cal S}$}]    Set of states of the RL algorithms
	\item [{\small${\cal A}$}]    Set of actions 
	\item [{\small${\cal R}$}]      Set of rewards
	\item [{\small${\cal T}$}]      Set of time steps
 \end{description}

\noindent \emph{Indexes}:
\begin{description}[itemsep=0.5ex, labelwidth=1.2cm, leftmargin=1.4cm]
	\item [{\small$i$}]  DG unit $i\in{\cal{G}}$
	\item[{\small$j$}] ESS $j\in\cal{B}$
	\item[{\small$m$}] PV unit $m\in\cal{V}$
	\item[{\small$k$}] Load demand $m\in\cal{L}$
	\item [{\small$t$}]  Time-step $t\in\cal{T}$
\end{description}

\noindent \emph{Parameters}:
\begin{description}[itemsep=0.5ex, labelwidth=1.2cm, leftmargin=1.4cm]
	\item [{\small$a_{i}~b_{i}~c_{i}$}]    	   	Quadratic, linear and constant parameters associated to the $i$-th DG operation cost
	\item [{\small$\Delta t$}]    		Length for the discretization of the operational time
	\item [{\small$\lambda$}]    		Discount factor 
	\item [{\small$\overline{P}_{t}^{G}$ $\underline{P}_i^{G}$}]    	Maximum and minimum generation limit of the DG units
	\item[\small$RU_{i}$~$RD_{i}$]	Ramping up and ramping down ability of the DG units 
    \item[{\small$\overline{P}_{j}^{B}$~$\underline{P}_{j}^{B}$}]     Maximum and minimum charging/discharging limit of the ESSs
    \item[{\small$\overline{E}_{j}^{B}$~$\underline{E}_{j}^{B}$}]     Maximum and minimum level of SOC of the ESSs
    \item[\small$\overline{P}^{C}$]
    Maximum main grid export/import limit
	\item [{\small$\beta$}]    			Electricity sell coefficient
	\item [{\small$\eta_{B}$}] Energy exchange efficiency for ESSs
	\item [{\small$\sigma_{1}$~$\sigma_{2}$}]   Reward re-scale and constrain penalty coefficients	
	\item[{\small$\rho_{t}$}] Electricity price for time slot $t$
	\item [{\small$P^{V}_{m,t}$}]    		    Active power of PV systems
	\item[{\small$P^{L}_{k,t}$}]	Active power demand 
	\end{description}
\noindent \emph{Continuous Variables}:
\begin{description}[itemsep=0.5ex, labelwidth=1.2cm, leftmargin=1.4cm]
	\item [{\small$P^{G}_{i,t}$}]    		Active power output of DG units
	\item [{\small$P^{B}_{j,t}$}]    			Active	power discharge/charge of ESSs
	\item [{\small$SOC_{j,t}^{B}$}]    			State of charge for ESSs
	\item[{\small$P_{t}^{N}$}]
		Active power exported/imported of main grid
	\item[{\small$\Delta P_{t}$}]	Active power unbalance 
\end{description}

\noindent \emph{Binary Variables}:
\begin{description}[itemsep=0.5ex, labelwidth=1.2cm, leftmargin=1.4cm]
	\item [{\small$u_{i,t}$}]    		Operational state of the DG units
\end{description}

\section{Introduction}
The massive integration of renewable-based resources inherently increases the complexity of the energy systems' scheduling problem~\cite{olivares_centralized_2011}. Existing scheduling approaches are not adequate to deal with a large number of distributed generation (DG) units and the increasing level of uncertainty. Innovative approaches are urgently needed, capable of providing in real-time good-quality solutions while still enforcing operational and technical constraints. In the technical literature, two main approaches are available to deal with the energy system's uncertainty, namely \textit{model-based} and \textit{model-free} approaches~\cite{zia_microgrids_2018}. In model-based approaches, uncertainty is considered either by using a probability distribution function or by leveraging a set of representative scenarios, constructing a complex mathematical formulation considering the system's operational constraints. This formulation leads to a stochastic multi-stage mathematical programming model that can be solved using commercial mixed-integer nonlinear programming (MINLP) solvers~\cite{hafiz2019energy,vergara_optimal_2019}, such as CPLEX. Nevertheless, although capable of providing good quality solutions, existing model-based approaches are not adequate for real-time operation due to the large computational time required.

To overcome this, model-free approaches have been introduced as an alternative solution. The most promising model-free approach is based on the use of reinforcement learning (RL)~\cite{chen_reinforcement_2021}. In this, by modeling the decision-making problem as a Markov Decision Process (MDP), RL algorithms can learn the system's dynamics by interaction, providing good-quality solutions guided by a reward value used as a performance indicator~\cite{sutton_reinforcement_2018}. The main advantage of RL-based algorithms is that, if properly designed and trained, they can capture energy systems' uncertainty by leveraging historical data, providing good-quality solutions in real-time, and avoiding any computational burden during operation.

Recently, Deep reinforcement learning (DRL), using Deep Neural Networks (DNN) as function approximators, has shown good performance on solving different energy-related scheduling problems with continuous actions, including home appliances~\cite{nakabi_deep_2021} and microgrids scheduling~\cite{ji_real-time_2019}. Unfortunately, these works rely on value-based RL algorithms, which do not scale well. Instead, policy-based DRL algorithms, by integrating an actor-critic structure, have shown outstanding performance on complex MDP tasks such as robot control and video games~\cite{schulman_high-dimensional_2018}. The current DRL state-of-the-art includes algorithms such as Proximal Policy Optimization (PPO)~\cite{schulman_proximal_2017}, Soft Actor-Critic (SAC), Policy optimization~\cite{haarnoja_soft_2018}, Deep Deterministic Policy Gradient (DDPG)~\cite{lillicrap_continuous_2019} and derivation (TD3)\cite{fujimoto_addressing_2018}.
 
Unlike general MDP tasks, the energy system optimal scheduling problem, formulated as an MDP, must enforce a rigorous set of operational constraints to ensure a reliable operation. For instance, the power balance constraint, modeled as an equality constraint, must always be met during real-time operation. Enforcing equality constraints in DRL algorithms is currently an unresolved challenge. Several works indirectly enforced the power balance constraint, for example, by defining a specific DG unit as an infinite source of power~\cite{vazquez-canteli_citylearn_nodate} or by adding a penalty term to the reward function~\cite{ji_data-driven_2021, zhou_combined_2020}. Nevertheless, this introduces extra hyperparameters that impact the DRL algorithms' performance and capability of providing feasible solutions. Given the challenge mentioned above, in this paper,  we aim to provide a fair performance comparison of DRL algorithms for energy systems optimal scheduling problems. To do this, we compared DDPG, TD3, SAC, and PPO algorithms' performance and performed a sensibility analysis of hyperparameters on the algorithms' capabilities of providing good-quality solutions, even in unseen operational scenarios. 

\vspace{-2mm}
\section{Mathematical Programming Formulation}\label{miqp_formulation}
The energy systems optimal scheduling problem can be modeled as the the MINLP model given by \eqref{eq_obj}-\eqref{eq_binary}. The objective function in \eqref{eq_obj} aims at minimizing the operational cost for the whole time horizon ${\cal T}$, which comprises the operational cost of the DG units, as presented in \eqref{eq_gen_cost} and the cost of buying/selling electricity from/to the main grid, as in \eqref{eq_power_exchange_cost}. Given the output power of DG units $P_{i,t}^{G}$, the operational cost can be estimated by  using a quadratic model as in \eqref{eq_gen_cost}. The transaction cost between the energy system and the main grid is settled according Time-of-Use (TOU) prices, in which it is assumed that selling prices are lower than the purchasing prices. In \eqref{eq_power_exchange_cost}, $\rho_{t}$ is the TOU price at time slot $t$, while $P_{t}^{N}$ refers to the exported/imported power transaction to/from the main grid.

\vspace{-2mm}
\begin{equation}\label{eq_obj}
\min\sum_{t\in\cal{T}}\sum_{i\in\cal{G}}(C_{i,t}^{G}+C_{t}^{E})\Delta t
\end{equation}

\vspace{-2mm}
\begin{equation}\label{eq_gen_cost}
C_{i,t}^{G}=a_{i}\cdot\left(P_{i,t}^{G}\right)^{2}+b_{i}\cdot P_{i,t}^{G}+c_{i},  \quad i \in \cal{G}.
\end{equation}

\vspace{-2mm}
\begin{equation}\label{eq_power_exchange_cost}
C_{t}^{E} = \begin{cases}
		\rho_{t}P_{t}^{N}	  &  \quad P_{t}^{N}>0,\\
		\beta\rho_t P_{t}^{N} & \quad P_{t}^{N}<0.\\
	\end{cases}
\end{equation}


Subject to: 
\vspace{-2mm}
\begin{multline}\label{eq_balance}
\sum_{i\in\cal{G}}P_{i,t}^{G} +\sum_{m\in\cal{V}}P_{m,t}^{V}+P_{t}^{N}+\sum_{j \in\cal{B}}P_{j,t}^{B} = \sum_{k \in \cal{L}}P_{k,t}^{L}, \forall t \in \cal{T}
\end{multline}

\vspace{-4mm}

\begin{flalign}
&\underline{P}^{G}_{i}\cdot u_{i,t} \leq P^{ G}_{i,t}\leq \overline{P}^{G}_{i}\cdot u_{i,t}
& \forall i \in {\cal G}, \forall t \in {\cal{T}} 
& \label{eq_max_min_output_constrain} \\
&P_{i,t}^{G}-P_{i,t-1}^{G}\leq RU_{i}
& \forall i \in {\cal G}, \forall t \in {\cal{T}} 
& \label{eq_ramping_up_constrain} \\
&P_{i,t}^{G}-P_{i,t+1}^{G}\leq RD_{i}
& \forall i \in {\cal G}, \forall t \in {\cal{T}} 
& \label{eq_ramping_down_constrain} \\
&-\underline{P}_{j}^{B}\leq P_{j,t}^{B}\leq \overline{P}_{j}^{B}
&  \forall j \in {\cal{B}}, \forall t \in {\cal{T}}
& \label{eq_char_disc_cons} \\
&SOC_{j,t}^{B}=SOC_{j,t-1}+\eta_{B}P_{j,t}^{B}\Delta t
&\forall j \in {\cal{B}}, \forall t \in {\cal{T}}
&\label{eq_SOC_cha} \\
&\underline{E}_{j}^{B}\leq SOC_{j,t}^{B}\leq\overline{E}_{j}^{B}
&\forall j \in {\cal{B}}, \forall t \in {\cal{T}}
&\label{eq_SOC_cons} \\
&-\overline{P}^{C}\leq P_{t}^{N}\leq \overline{P}^{C}
&\forall t \in {\cal{T}}
& \label{eq_export_cons} \\
& u_{i,t}\in \{0,1\}
&\forall i \in {\cal{G}}, \forall t\in {\cal{T}}
& \label{eq_binary}
\end{flalign}





Expression \eqref{eq_balance} defines the power balance constrain. A binary variable is used to model the DG unit's commitment status, i.e., $u_{i,t}=1$ represent that the $i{\text{-th}}$ DG unit is operating. Expression \eqref{eq_max_min_output_constrain} defines the DG units generation power limits while \eqref{eq_ramping_up_constrain} and \eqref{eq_ramping_down_constrain} enforce the DG unit's ramping up and down constraints, respectively. Energy storage systems (ESSs) are modeled using \eqref{eq_char_disc_cons}-\eqref{eq_SOC_cons}. In this model, the operation cost of ESSs is not considered, while ESSs are allowed to schedule their discharge and charge power in advance. Expression \eqref{eq_char_disc_cons} defines the charging and discharging power limits, while expression \eqref{eq_SOC_cha} models the state of charge (SOC) as a function of the charging and discharging power. Expression in \eqref{eq_SOC_cons} limits the energy stored in the ESSs, avoiding the impacts caused by over-charging and over-discharging. Finally, main grid export/import power limit is modeled by the expression in \eqref{eq_export_cons}. 

\vspace{-2mm}
\section{MDP Formulation}\label{mdp_formulation}
The above-presented problem can be formulated as a MDP, which can be represented as a 5-tuple $(\cal{S},\cal{A},\cal{P},\cal{R},\lambda)$, where $\cal{S}$ represents the set of system states, $\cal{A}$ the set of actions, $\cal{P}$ the state transition probability, $\cal{R}$ the reward function, and $\lambda$ the discount factor. In this formulation, the energy system operator can be modeled as an RL agent. The state information provides an important basis for the operator to dispatch units. We define a state as $s_{t}= (P_{t}^{V},P_{t}^{L},P_{t-1}^{G},SOC_{t}), \quad s_{t}\in \mathcal{S}$, while the actions, defining the scheduling of the DG units and the ESSs, as $
a_{t}= (P^{G}_{i,t},P_{t}^{B}), \quad a_{t}\in \mathcal{A}$. Notice that the RL agent does not directly control the transaction between the energy system and the main grid. Instead, after any action is executed, power is exported/imported from the main grid to maintain the power balance. Nevertheless, a maximum power capacity constraint exits and must be enforced i.e.,~\eqref{eq_export_cons}.

Given the state $s_{t}$ and action $a_{t}$ at time step $t$, the energy system transit to the next state $s_{t+1}$ defined by
\vspace{-2mm}
\begin{equation}
\mathcal{P}_{s s^{\prime}}^{a}=\operatorname{Pr}\left\{s_{t+1}=s^{\prime} \mid s_{t}=s, a_{t}=a\right\}
\end{equation}
where $\mathcal{P}_{ss'}^{a}$ corresponds to the transition probability, which models the energy system's dynamics and uncertainty involved. In model-based algorithms, the uncertainty is predicted by a determined value or sampling from a prior probability distribution. Instead, DRL is a model-free approach, capable of learning the uncertainty and dynamics from historical data and interactions.

The reward $r_{t}$ is given by the environment as an indicator to guide update direction of the policy. In the energy systems optimal scheduling problem, the reward function should guide the RL agent to take actions that minimize the operational cost while enforcing the power balance constraint. This can be done by using the next reward function

\vspace{-2mm}
\begin{equation}
\label{eq_reward}
r_{t}\left(s_{t}, a_{t}\right)=-\sigma_{1}\left[\sum_{i\in\cal{G}} C^{G}_{i,t}+C^{E}_{t}\right] -\sigma_{2} \Delta P_t,
\end{equation}
in which $\Delta P$ corresponds to the power unbalance at time-step $t$ and is defined as,

\begin{equation}\label{eq_unb}
\Delta P_t=\lvert \sum_{i\in\cal{G}}P_{i,t}^{G}+\sum_{m\in\cal{V}}P_{m,t}^{V}+P_{t}^{N}+\sum_{j \in\cal{B}}P_{j,t}^{B}-\sum_{k \in \cal{L}}P_{k,t}^{L}\rvert
\end{equation}
while $\sigma_2$ is used to control the trade-off between the cost minimization and the penalty incurred in case of power unbalance. The goal of the RL algorithms is to find an optimal scheduling policy $\pi^*$(stochastic or deterministic) to maximize the total expected discounted rewards for the formulated MDP
\vspace{-2mm}
\begin{equation}
\pi^{*}=\arg \max _{\pi} \mathbb{E}_{\left(s_{t}, a_{t}\right) \sim \rho_{\pi}}\left[\sum_{t\in\cal{T}} R\left(s_{t}, a_{t}\right)\right].
\end{equation}

\section{DRL Policy-Based Algorithms}
The formulated MDP have a continuous state and action spaces, which can be hard to solve by using classical RL algorithms, such as $Q$ learning, due to their poor scalability features~\cite{sutton_reinforcement_2018}. By leveraging the generalization and fitting capability of DNNs, DRL algorithms have shown good performance when dealing with this challenge. In valued-based DRL algorithms, the action-state function $Q$ is iteratively updated to indirectly define a deterministic policy, for which the foundation is Bellman optimality equation:
\vspace{-2mm}
\begin{equation}
Q^{*}(s, a)=R(s, a)+\gamma \sum_{s^{\prime} \in \mathcal{S}} \mathcal{P}\left(s^{\prime} \mid s, a\right) \max _{a^{\prime} \in \mathcal{A}} Q^{*}\left(s^{\prime}, a^{\prime}\right)
\end{equation}

If the optimal value function $Q^{*}(s,a)$ is estimated, then the optimal policy can be derived as $\pi^{*}(s)= \mathop{\arg\max}_{a\in \mathcal{A}}Q^{*}(s,a)$. Value-based DRL algorithms use DNNs approximate the $Q$-function, dealing with continuous state spaces. However, for continuous action MDP problems, it requires a full scan of the action space when executing policy improvement i.e., $\arg \max _{a \in \mathcal{A}} Q^{\pi}(s, a)$, leading to a dimensionality problem. Instead, policy-based DRL algorithms directly search for the optimal policy. The policy is usually modeled with a parametric function, denoted as $\pi_{\theta}(s|a)$. Based on the policy gradient theorem, the policy gradient is expressed as $\nabla_{\theta} J(\theta)=\mathbb{E}_{\pi}\left[Q^{\pi}(s, a) \nabla_{\theta} \ln \pi_{\theta}(a \mid s)\right]$, where $J(\theta)$ is the reward function. Using gradient ascent, $\theta$ can be updated towards the direction suggested by  $\nabla_{\theta}J(\theta)$ to find the policy $\pi_{\theta}$ that lead the highest total discounted rewards. 

In this paper, we will compare the performance of state-of-the-art DRL algorithms, including DDPG, TD3, SAC, PPO when solving the MDP described in Sec.~\ref{mdp_formulation}, using a penalty to enforce the power balance constrain, as showed in \eqref{eq_reward}. DDPG and TD3 are off-policy, deterministic algorithms, while PPO and SAC are stochastic algorithms. In general, DRL algorithms interact with the environment to collect a reward. This reward is then used to update a critic DNNs parameters based on the temporal difference (TD) algorithm. Then, the critic network is used to update actor DNNs parameters based on policy gradient theory. A more detailed explanation of policy-based algorithms can be found in~\cite{chen_reinforcement_2021}.


\section{Experimental results}

\subsection{Experimental Setting}
Three DG units, using the parameters shown in Table~\ref{tab:dg_data_num}, are defined. For the ESSs, the charging and discharging limits are set as 100 kW, the nominal capacity as 500 kW, while the energy efficiency is set as $\eta_{B}=0.9$. We assume that the grid's maximum export/import limit is defined as 100 kW. To encourage the use of renewable energies, we set selling prices as half of the current electricity prices, i.e., $\beta=0.5$. One-hour resolution data profiles including solar generation, demand consumption, and electricity prices for a period of one year are used. 
The original data is divided into training and testing sets. The training set contains the first three weeks of each month, while the testing sets contain the remaining data. This strategy allows the DRL algorithm to consider seasonal changes in the PV generation and demand consumption. During training, the  EESs' initial SOC were randomly set. All algorithms were implemented in Python using PyTorch and trained for 1000 episodes. Unless otherwise mentioned, default settings were used. The implemented DRL algorithms and the environment are openly available at \cite{Shengren}. Hyperparameters $\sigma_1$ and $\sigma_2$ were defined as 0.01 and 50, respectively, as default values. Each experiment is run with five random seeds to eliminate randomness from the code implementation part during the simulation process. To evaluate the DRL algorithms' performance, the total operational cost, as in \eqref{eq_obj}, and the power unbalance as in \eqref{eq_unb}, are used as metrics. To validate and compare the performance of the tested DRL algorithms, we have also solved the MIQP formulation presented in Sec.~\ref{miqp_formulation} using Pyomo\cite{hart_pyomo_2017}. 

\begin{table}[t]
	\centering
	\caption{DG units information} \vspace{-2mm}
	\scalebox{0.8}{
		\begin{tabular}{cccccccc}
			\toprule
		Units & $a$[$\text{\$/kW}^2$] & $b$[\$/kW] & $c$[\$] & $\underline{P}^{G}$[kW] & $\overline{P}^{G}$[kW]& $RU$[kW]&$RD$[kW]\\
			\midrule
			$DG_1$ & 0.0034 & 3 & 30 & 10 & 150&100&100  \\
			\midrule
			$DG_2$ & 0.001 & 10 & 40 & 50 & 375&100&100  \\
			\midrule
			$DG_3$ & 0.001 & 15 & 70 & 100 & 500&200&200  \\
            \bottomrule
	\end{tabular}}%
	\label{tab:dg_data_num}%
	\vspace{-2mm}
\end{table}%



\subsection{Performance on the Training Set}
\begin{figure*}[h]
    \centering
    \includegraphics[width=1.8\columnwidth]{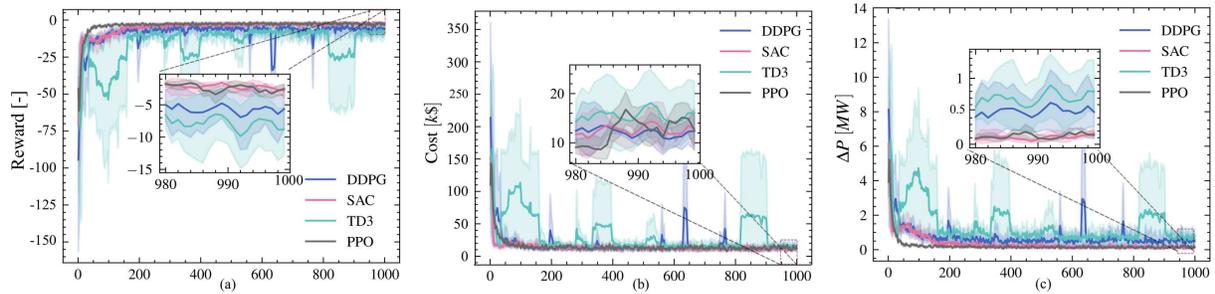}
        \vspace{-4mm}
    \caption{Average (a) reward, (b) operational cost, and (c) power unbalance, during 1000 episode training for all the tested DRL algorithms.}
    \label{fig_case1_loss_reward_cost}
\end{figure*}

Fig. \ref{fig_case1_loss_reward_cost} shows the average reward, losses, operational cost, and power unbalance for all the DRL algorithms during the training process. Reward increased rapidly after training for 100 episodes, while loss decreased to a small value after 200 episodes. This trend is due to the fact that the parameters of all algorithms are initialized randomly, leading to random actions. As a result, a high power unbalance is observed. As the training continues, all DRL agents learn how to meet the power unbalance constraint due to the penalty term used in the reward function. The comparison of the four algorithms shows that the PPO and SAC algorithms are more stable when compared with the DDPG and TD3 algorithms. Not only do PPO and SAC algorithms have a higher reward after training, but they also have a lower variance. In the end, all algorithms converged to similar values. However, DDPG and TD3 algorithms showed a significant performance decrease after training for 800 episodes. For PPO and SAC algorithms, the power unbalance is mitigated notably after training for 100 episodes and nearly disappeared at 1000 episodes, while DDPG and TD3 algorithms show that power unbalances decrease notably after 200 episodes and then remain after training 1000 episodes, at around 500-700 kW. 

\begin{figure}[t]
    \centering
    \subfigure[]{\includegraphics[width=0.45\columnwidth]{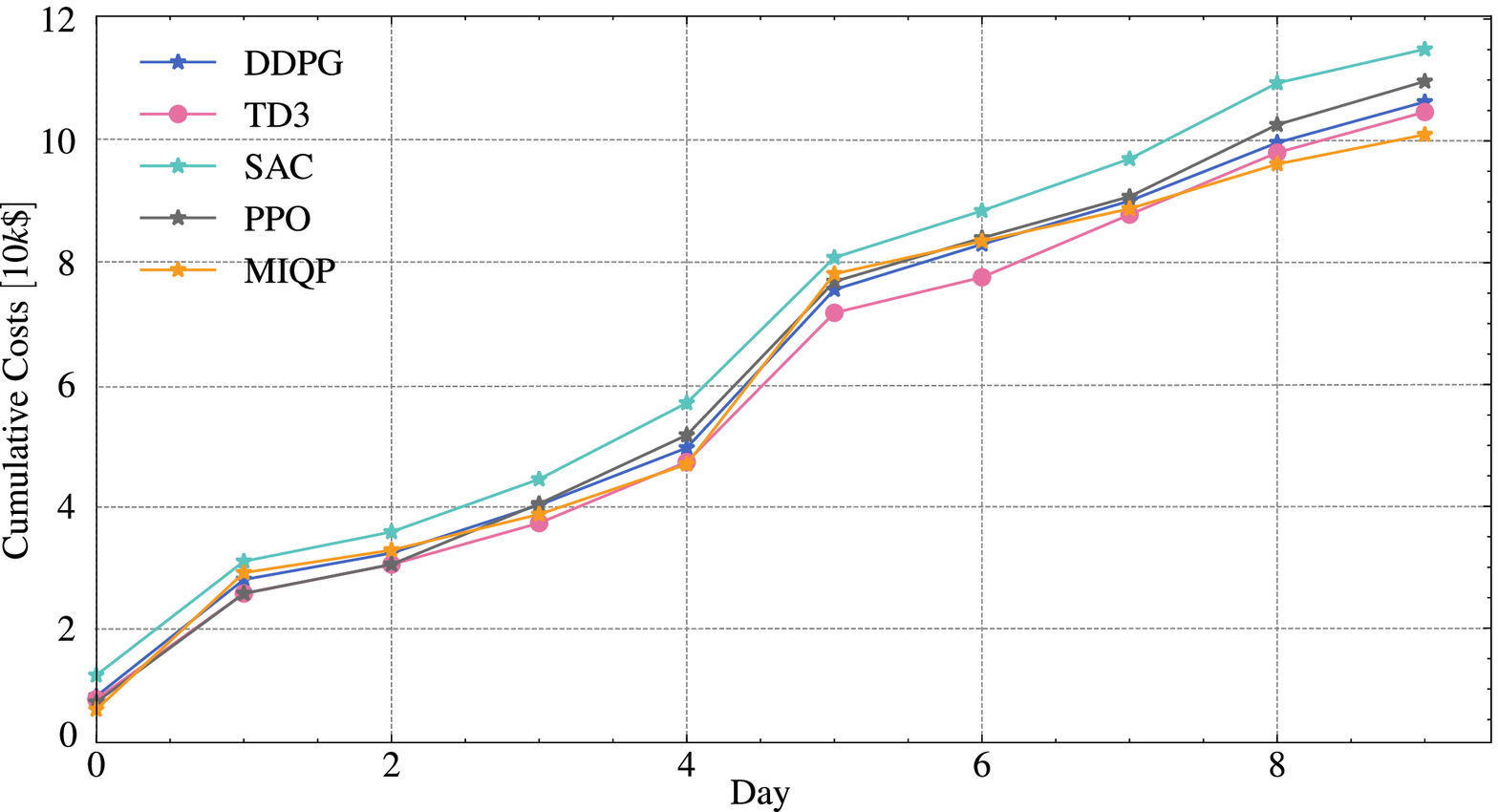}
    \label{cumulative_cost}}
    \subfigure[]{\includegraphics[width=0.45\columnwidth]{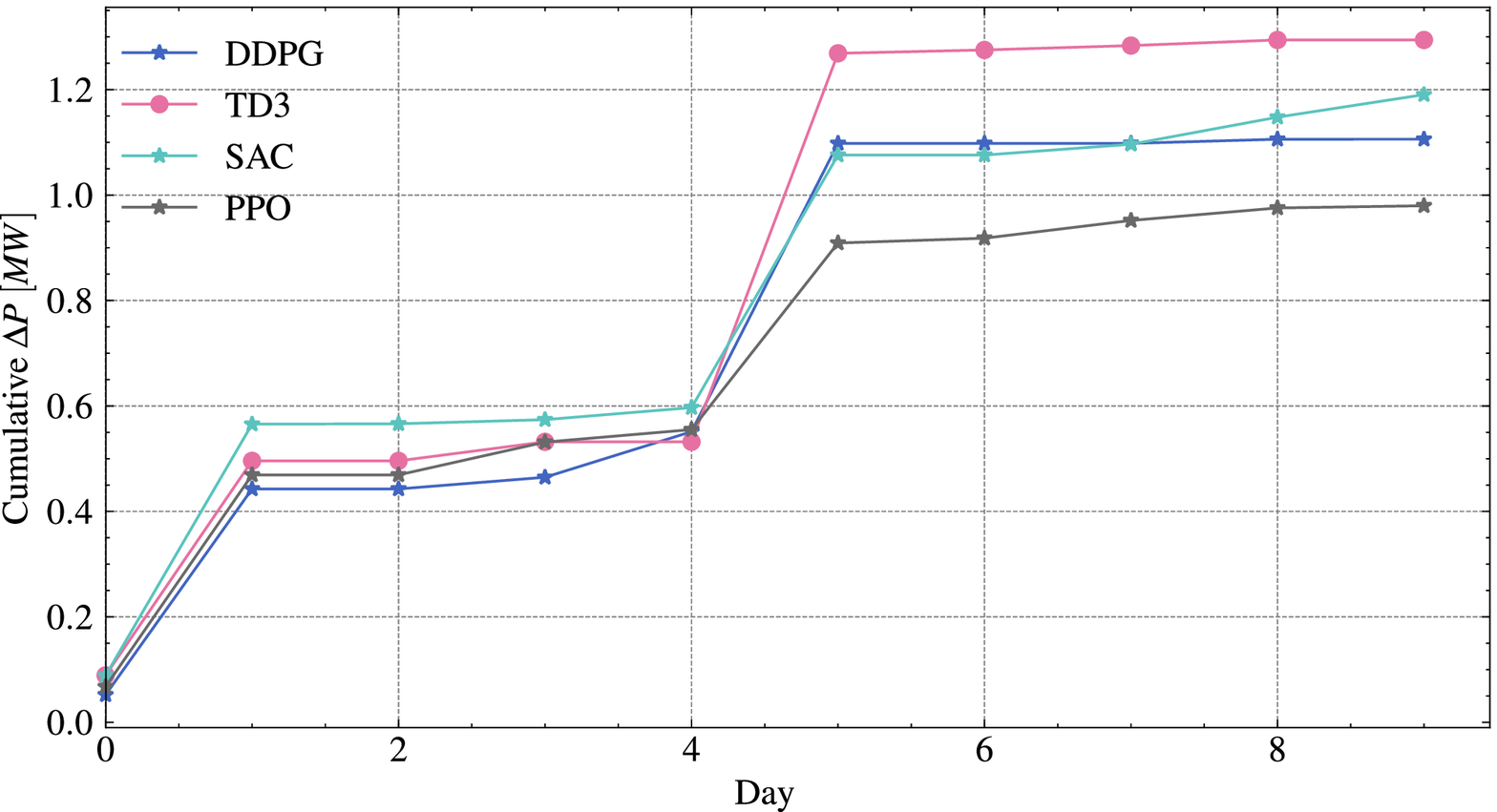}
    \label{cumulative_unbalance}}
    \caption{Cumulative operational cost and power unbalance for 10 days}
\end{figure}

Fig.~\ref{cumulative_cost} shows the cumulative operational cost for 10 days in the training set. All tested DRL algorithms have a similar performance compared to MIQP. The solution provided by the MIQP formulation is treated as the optimal solution. Notice that some of the DRL algorithms have outperformed the solution provided by the MIQP formulation in some days in terms of the operational cost. Nevertheless, this is because they did not strictly meet the power balance constraint. Fig.~\ref{cumulative_unbalance} shows that the cumulative power unbalances of 10 days is less than 1200 kW, accounting for no more than 5\% of the total demand. Notice that the PPO and the TD3 algorithms maintained the lowest and largest power unbalance, respectively, among all algorithms.



\subsection{Performance on the Test Set}

Table~\ref{tab_cum_cost} and Table~\ref{tab_cum_unb} shows the operational cost and the power unbalance, respectively, with 95\% confidence intervals, tested on 10 days on the test set. As can be seen, DRL algorithms show similar performance compared with the optimal solutions provided by the MIQP formulation. Among these, the PPO and TD3 algorithms over-perform the DDPG and SAC algorithms regarding the operational cost. Notice that all DRL algorithms were able to operate with a low power unbalance on most of the days from the test set. Nevertheless, all these algorithms failed to maintain the power unbalance in the last test day due to the significantly high demand consumption around peak time. This result shows that in the case of extreme events (e.g., high demand consumption, low renewable generation, etc.), perhaps not captured on the trained data, DRL algorithms fail to provide a feasible solution. These results are important as larger power unbalances can lead to an outage, especially if an export limit is defined with the main grid.  

\begin{table*}[t]
\caption{Mean and 95{\%} confidence bounds for cost$[\$]$}\vspace{-2mm}
\centering
\scalebox{0.9}{
\begin{tabular}{cccccc}
\hline
\textbf{Test day} & \textbf{DDPG}                 & \textbf{TD3}                & \textbf{SAC}               & \textbf{PPO}                  & \multicolumn{1}{l}{\textbf{MIQP}} \\ \hline
\textbf{1}        & 21142 (20670, 21613)          & 19810 (19332, 20287)        & 21907 (21445, 22368)       & \textbf{17637 (14385, 20888)} & 18145                             \\
\textbf{2}        & 11364 (10784, 11943)          & \textbf{9824 (9306, 10342)} & 11084 (8246, 13921)        & 10833 (9714, 11924)           & 8358                              \\
\textbf{3}        & 9539 (9162, 9915)             & \textbf{8868 (8442, 9293)}  & 10362 (8932, 11792)        & 10214 (8980, 11446)           & 7417                              \\
\textbf{4}        & 3292 (1886, 4697)             & 5120 (4742, 5496)           & 2502 (1469, 3534)          & \textbf{2478 (1947, 3009)}    & 1046                              \\
\textbf{5}        & 23128 (21024, 25232)          & 23038 (22247, 23829)        & 20146 (17179, 23112)       & \textbf{19045 (17399, 20690)} & 14971                             \\
\textbf{6}        & \textbf{11266 (10774, 11758)} & 15631 (15169, 16092)        & 13345 (12197, 14491)       & 12844 (11018, 14669)          & 8085                              \\
\textbf{7}        & 4046 (2608, 5484)             & 4901 (4635, 5166)           & \textbf{2766 (1636, 3894)} & 3076  (2254, 3839)            & 1652                              \\
\textbf{8}        & 11219 (10345, 12092)          & \textbf{9761 (9424, 10097)} & 12019 (8803, 15233)        & 10778 (8762, 12792)           & 7126                              \\
\textbf{9}        & 7349 (5838, 8860)             & 7662 (7190, 8134)           & 7943 (6017, 9869)          & \textbf{7250 (5881, 8619)}    & 5053                              \\
\textbf{10}       & 26285 (24625, 27944)          & 24050 (23857, 24242)        & 23303 (22167, 24439)       & \textbf{22614 (20087, 25141)} & 23452                             \\ \hline
\label{tab_cum_cost}
\end{tabular}}
\vspace{-2mm}
\end{table*}

\begin{table*}[t]
\caption{Mean and 95{\%} confidence bounds for unbalance power [kW]}\vspace{-2mm}
\centering
\scalebox{0.9}{
\begin{tabular}{ccccc}
\hline
\textbf{Test day} & \textbf{DDPG}                & \textbf{TD3}            & \textbf{SAC}              & \textbf{PPO}                     \\ \hline
\textbf{1}        & 189.89 (189.65, 190.14)      & 246.14 (245.52, 246.76) & 225.56 (224.08, 227.04)   & \textbf{171.27 (29.19, 313.35)}  \\
\textbf{2}        & \textbf{5.45 (0, 23.46)}     & 12.63 (12.63, 12.63)    & 8.30 (0, 39.89)           & 27.94 (0, 66.88)                 \\
\textbf{3}        & 74.16 (34.41, 113.92)        & 49.43 (49.43, 49.43)    & \textbf{25.73 (0, 51.79)} & 36.12 (0, 110.21)                \\
\textbf{4}        & 2.27 (0, 12.64)              & 37.42 (31.28, 43.57)    & \textbf{0.01 (0, 0.03)}   & 36.70 (0.75, 72.65)              \\
\textbf{5}        & \textbf{10.85 (8.31, 13.39)} & 73.40 (73.18, 73.62)    & 100.75 (70.38, 131.11)    & 67.81 (0, 185.54)                \\
\textbf{6}        & 32.34 (32.34, 32.34)         & \textbf{0.67 (0, 3.59)} & 50.05 (35.37, 64.73)      & 20.52 (0, 84.23)                 \\
\textbf{7}        & 0.07 (0, 0.48)               & \textbf{0.00 (0, 0)}    & 37.72 (37.72, 37.72)      & 3.38 (0, 13.12)                  \\
\textbf{8}        & 64.61 (33.82, 95.41)         & 107.24 (107.24, 107.24) & 32.38 (22.84, 41.91)      & 66.00 (14.92, 117.08)            \\
\textbf{9}        & 33.88 (10.52, 57.24)         & 19.81 (19.74, 19.89)    & \textbf{0.09 (0, 0.63)}   & 28.51 (0, 78.86)                 \\
\textbf{10}       & 517.19 (506.65, 527.74)      & 436.86 (433.39, 440.33) & 450.06 (434.73, 465.39)   & \textbf{410.11 (331.57, 488.66)} \\ \hline
\label{tab_cum_unb}
\end{tabular}}
\vspace{-2mm}
\end{table*}

\begin{figure*}[t]
    \centering
    \includegraphics[width=2\columnwidth]{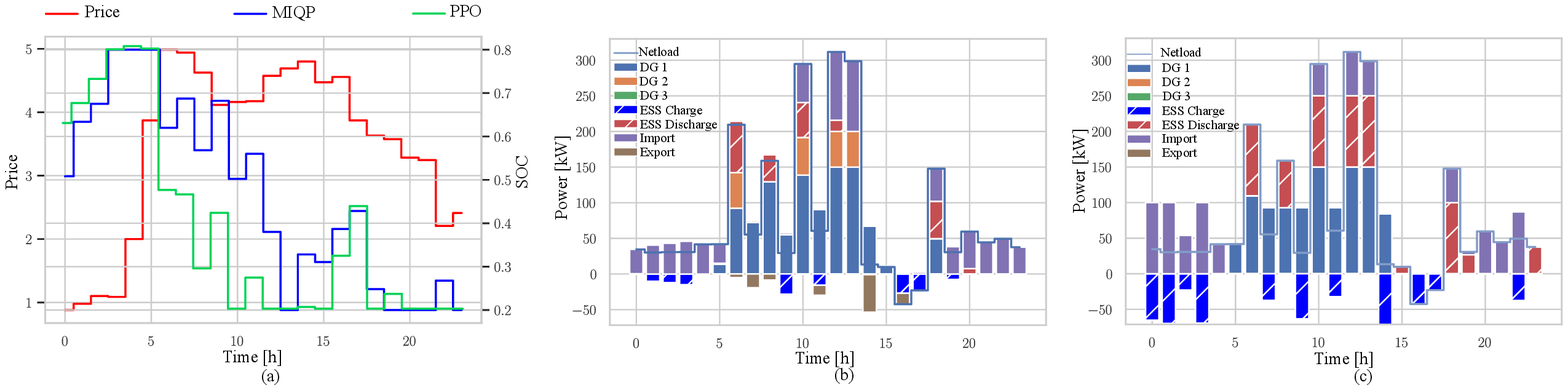}
    \vspace{-2mm}
    \caption{Operational schedule of all DG units and ESSs: $(a)$ and $(b)$ PPO algorithm,  $(b)$ and $(c)$ MIQP formulation.}
    \label{optimal_schedules}
    \vspace{-2mm}
\end{figure*}

Fig.~\ref{optimal_schedules} show the ESSs' SOC and the output power of the DG units, ESSs, and the export/import power from the grid defined using the PPO algorithm and the optimal solution provided by the MIQP formulation. As can be seen, when the electricity price is high and the net power is low, the PPO algorithm dispatches the ESSs in charging mode, while around peak-time, the ESSs is dispatched in discharging mode. A similar operational schedule is defined by the optimal solution provided by the MIQP formulation. As can be seen in Fig.~\ref{optimal_schedules}, at 12h, the DG unit 1 and the ESSs operating in discharge mode play the most critical role in meeting all the demand. A different operational schedule was defined by the PPO algorithms for the same time step, in which the DG unit 2 and the main grid supply all the demand consumption. In this regard, as the schedule defined by the PPO algorithm did not prioritize the use of the ESSs, a higher operational cost was observed. 

\subsection{Sensitivity Analysis}
Enforcing technical constraints using a penalty term in the reward function is one of the most common approaches used to provide feasible solutions. As previously mentioned, this approach introduces extra hyperparameters, in this case, to balance the numeric relationship between the operational cost and the power balance constraint. Additionally, as the grid export/import power limit provide flexibility for the actions taken by DRL algorithms, this can significantly impact the feasibility of the provided solutions. In this section, a sensitivity analysis of the penalty coefficient, $\sigma_2$, as used in expression \eqref{eq_reward}, is presented. 

\begin{figure*}
    \centering
    \psfrag{a}[][][0.7]{$\rho_{2}=20$}
    \psfrag{b}[][][0.7]{$\rho_{2}=50$}
    \psfrag{c}[][][0.7]{$\rho_{2}=100$}
    \includegraphics[width=1.8\columnwidth]{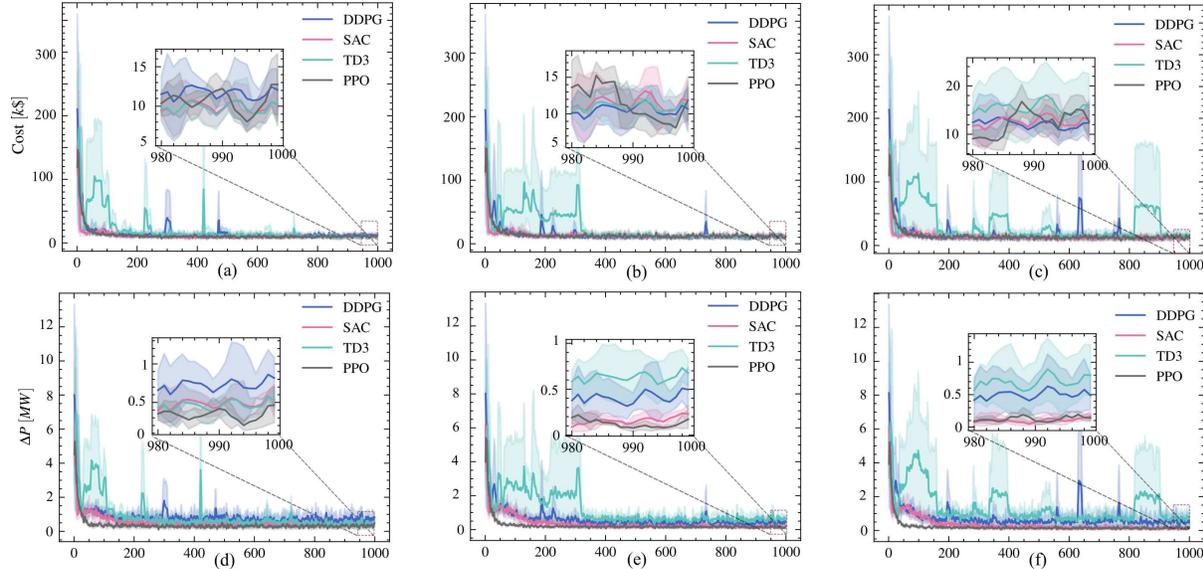}
    \vspace{-4mm}
    \caption{Sensitive analysis for $\sigma_{2}=(20,50,100)$, (a, b, c) and (d, e, f) refer to the cost and power unbalance corresponding to 20, 50, 100, respectively.}
    \label{fig_penalty_analysis}
\end{figure*}

Fig.~\ref{fig_penalty_analysis} shows the convergence performance of all the tested DRL algorithms for different values of the coefficient $\sigma_2$ considering $\overline{P}^C=100$~kW. As can be seen, the operational cost increased significantly when increasing $\sigma_2$ from 20 to 50 and then reduced after increasing it from 50 to 100. On the other hand, as expected, increasing $\sigma_2$ notably reduced the power unbalance. In this case, if $\sigma_2$ is set as 100, the PPO and SAC algorithms were able to almost completely mitigate the power unbalance. Nevertheless, the DDPG and TD3 algorithms were able to minimize the power unbalance for all values of $\sigma_2$, which allows concluding that such algorithms are less sensitive to this hyperparameter when compared with the PPO and SAC algorithms. 


\section{Conclusion}
In this paper, the performance of policy-based DRL algorithms (DDPG, TD3, SAC, PPO) has been evaluated for the energy systems optimal scheduling problem. We provided a detailed sensitivity analysis for the penalty terms used to enforce the power balance constraint and minimize the operational cost in the reward function definition. Results showed that DRL algorithms are able to provide good quality solutions in real-time by exploiting the good generalization capabilities of deep learning models. In general, we observed that the PPO algorithm outperforms DDPG, SAC, and TD3 algorithms. The penalty coefficient used to balance the relationship between the operational cost and enforce the power balance constraint significantly impacted the performance of all DRL algorithms. Additionally, we observed that increasing the grid export/import power limit increases the flexibility of actions taken by algorithms (capability of providing feasible actions solutions). Nevertheless, all the tested DRL algorithms lack safety guarantees compared to model-based approaches, which could impede their practical implementation. In this regard, new approaches to enforce operational constraints in DRL algorithms are urgently needed.   

\section*{Acknowledgment}
This work made use of the Dutch national e-infrastructure with the support of
the SURF Cooperative (grant no. EINF-2684) and was supported by the Chinese Scholarship Council (CSC) (grant no. 202106660002).
\vspace{0mm}

\bibliographystyle{IEEEtran} 
\vspace{-2mm}

\end{document}